\documentclass[journal]{IEEEtran}
\usepackage{amsmath,amsfonts}
\usepackage{algorithmic}
\usepackage{algorithm}
\usepackage{array}
\usepackage[caption=false,font=normalsize,labelfont=sf,textfont=sf]{subfig}
\usepackage{textcomp}
\usepackage{stfloats}
\usepackage{url}
\usepackage{verbatim}
\usepackage{graphicx}
\usepackage{cite}
\usepackage{amssymb}
\begin{document}

\title{\huge Two-Timescale Learning for Pilot-Free ISAC Systems}

\author{Jian Xiao, Ji Wang,~\IEEEmembership{Senior Member,~IEEE}, Qimei Cui,~\IEEEmembership{Senior Member,~IEEE}, Lihua Li, \IEEEmembership{Member,~IEEE}, Xingwang Li,~\IEEEmembership{Senior Member,~IEEE},  Yingzhuang Liu, and Tony Q. S. Quek,~\IEEEmembership{Fellow,~IEEE}
\thanks{
\emph{Corresponding author: Ji Wang.}

Jian Xiao and Ji Wang are with the Department of Electronics and Information Engineering, College of Physical Science and Technology, Central China Normal University, Wuhan 430079, China (e-mail: jianx@mails.ccnu.edu.cn; jiwang@ccnu.edu.cn).

Qimei Cui is with National Engineering Research Center for Mobile Network Technologies, Beijing University of Posts and Telecommunications, Beijing, 100876, China (e-mail: cuiqimei@bupt.edu.cn).

 Lihua Li is with the State Key Laboratory of Networking and Switching Technology, Beijing University of Posts and Telecommunications, Beijing 100876, China (e-mail: lilihua@bupt.edu.cn).

 Xingwang Li is with the School of Physics and Electronic Information Engineering, Henan Polytechnic University, Jiaozuo 454003, China (e-mail:
 lixingwang@hpu.edu.cn).
 
 Yingzhuang Liu is with the School of Electronic Information and Communications, Huazhong University of Science and Technology, Wuhan 430074, China (e-mail: liuyz@hust.edu.cn).
 
 Tony Q. S. Quek is with the Singapore University of Technology and Design, Singapore 487372, and also with the Department of Electronic Engineering, Kyung Hee University, Yongin 17104, South Korea (e-mail:tonyquek@sutd.edu.sg).
}}



\maketitle

\begin{abstract}
A pilot-free integrated sensing and communication (ISAC) system is investigated, in which phase-modulated continuous wave (PMCW) and non-orthogonal multiple access (NOMA) waveforms are co-designed to achieve simultaneous target sensing and data transmission. To enhance effective data throughput (i.e., Goodput) in PMCW-NOMA ISAC systems, we propose a deep learning-based receiver architecture, termed two-timescale Transformer (T3former), which leverages a Transformer architecture to perform joint channel estimation and multi-user signal detection without the need for dedicated pilot signals. By treating the deterministic structure of the PMCW waveform as an implicit pilot, the proposed T3former eliminates the overhead associated with traditional pilot-based methods. The proposed T3former processes the received PMCW-NOMA signals on two distinct timescales, where a fine-grained attention mechanism captures local features across the fast-time dimension, while a coarse-grained mechanism aggregates global spatio-temporal dependencies of the slow-time dimension. Numerical results demonstrate that the proposed T3former significantly outperforms traditional successive interference cancellation (SIC) receivers, which avoids inherent error propagation in SIC. Specifically, the proposed T3former achieves a substantially lower bit error rate and a higher Goodput, approaching the theoretical maximum capacity of a pilot-free system. 
\end{abstract}


\section{Introduction}
\IEEEPARstart{T}{he} relentless growth of wireless services, from Internet of Things (IoT) to autonomous vehicles, coupled with the bandwidth demands of high-resolution radar, has led to a contested RF spectrum \cite{ref1}. Integrated Sensing and Communication (ISAC) has emerged as a transformative solution, coalescing these dual functions to achieve significant gains in spectral efficiency, hardware cost, and power consumption \cite{ref2}.
Among various ISAC waveforms, the phase-modulated continuous wave (PMCW) waveform, widely used in automotive radars, presents a compelling case for integration \cite{Oliveira}. PMCW signals, typically based on pseudo-random binary sequences (PRBS), exhibit a sharp, thumbtack-like ambiguity function, which is ideal for achieving high-resolution range and Doppler measurements with minimal ambiguity. Furthermore, the phase of the waveform can be naturally modulated to carry communication data, making it a suitable candidate for a dual-function waveform. However, a primary drawback of conventional PMCW-based communication is its inherently low data rate. To ensure robust sensing performance, PMCW frames often involve repetitions of the same code sequence, with communication symbols modulating entire sequences or blocks of sequences. This structure, while beneficial for sensing, constrains the communication throughput.

To overcome this limitation, in this work, two synergistic techniques, i.e., multiple-input multiple-output (MIMO) and non-orthogonal multiple access (NOMA), are explored to improve communication throughout. MIMO technology enhances both sensing resolution via a larger virtual aperture and communication capacity via spatial multiplexing \cite{ref4}. Additionally, NOMA allows multiple users to be served within the same time-frequency resource block by allocating different power levels, thereby significantly boosting spectral efficiency \cite{ref5}. The combination of MIMO and NOMA in an ISAC context offers a promising pathway to serve multiple communication users at high data rates while simultaneously performing high-resolution sensing.
A fundamental challenge in ISAC system is obtaining accurate channel state information \cite{Gupta}. Traditional methods rely on transmitting dedicated pilot signals, which consume valuable time-frequency resources and thus reduce effective data throughput, i.e., Goodput \cite{QiaoGoodput}. This observation motivates the central question of this work: \emph{Can we design a high-performance ISAC system that dispenses with pilots?}

In this paper, we propose a pilot-free PMCW-NOMA ISAC system, which leverages the deterministic and known structure of the underlying code sequence of the PMCW waveform as an implicit pilot. This allows us to perform joint channel estimation and multi-user signal detection without overhead. However, the absence of explicit pilots coupled with the intricate interference structure of a PMCW-NOMA ISAC system, e.g., inter-user interference and multi-path effects, renders conventional receivers based on suboptimal zero-forcing or successive interference cancellation (SIC) with error propagation. Consequently, we propose a deep learning-based receiver architecture for the considered pilot-free ISAC system, termed two-timescale Transformer (T3former). The proposed T3former is specifically designed to interpret the complex spatio-temporal structure of the received signals, which operates on two timescales of PMCW-NOMA waveforms. Specifically, a fine-grained attention mechanism processes the local fast-time structure of the signal, while a coarse-grained attention mechanism learns the global dependencies across the entire slow-time signal block and antenna array. This hierarchical processing enables the model to implicitly learn the channel characteristics and effectively decode the NOMA signals for multiple users. Numerical results demonstrate that the proposed T3former significantly surpasses traditional receivers in terms of bit error rate (BER) and Goodput, while preserving the high-fidelity multi-target sensing capabilities.



\section{System Model}
We consider a monostatic ISAC system where a dual-functional base station (DF-BS) is equipped with a uniform linear array (ULA). The DF-BS aims to detect and estimate the parameters of $Q$ targets while simultaneously serving $K$ downlink NOMA users. The ULA consists of $N_t$ transmit antennas and $N_r$ receive antennas, with an inter-element spacing of $d = \lambda/2$, where $\lambda$ denotes the carrier wavelength.

\begin{figure}[!t]
\centering
\includegraphics[width=2.7in]{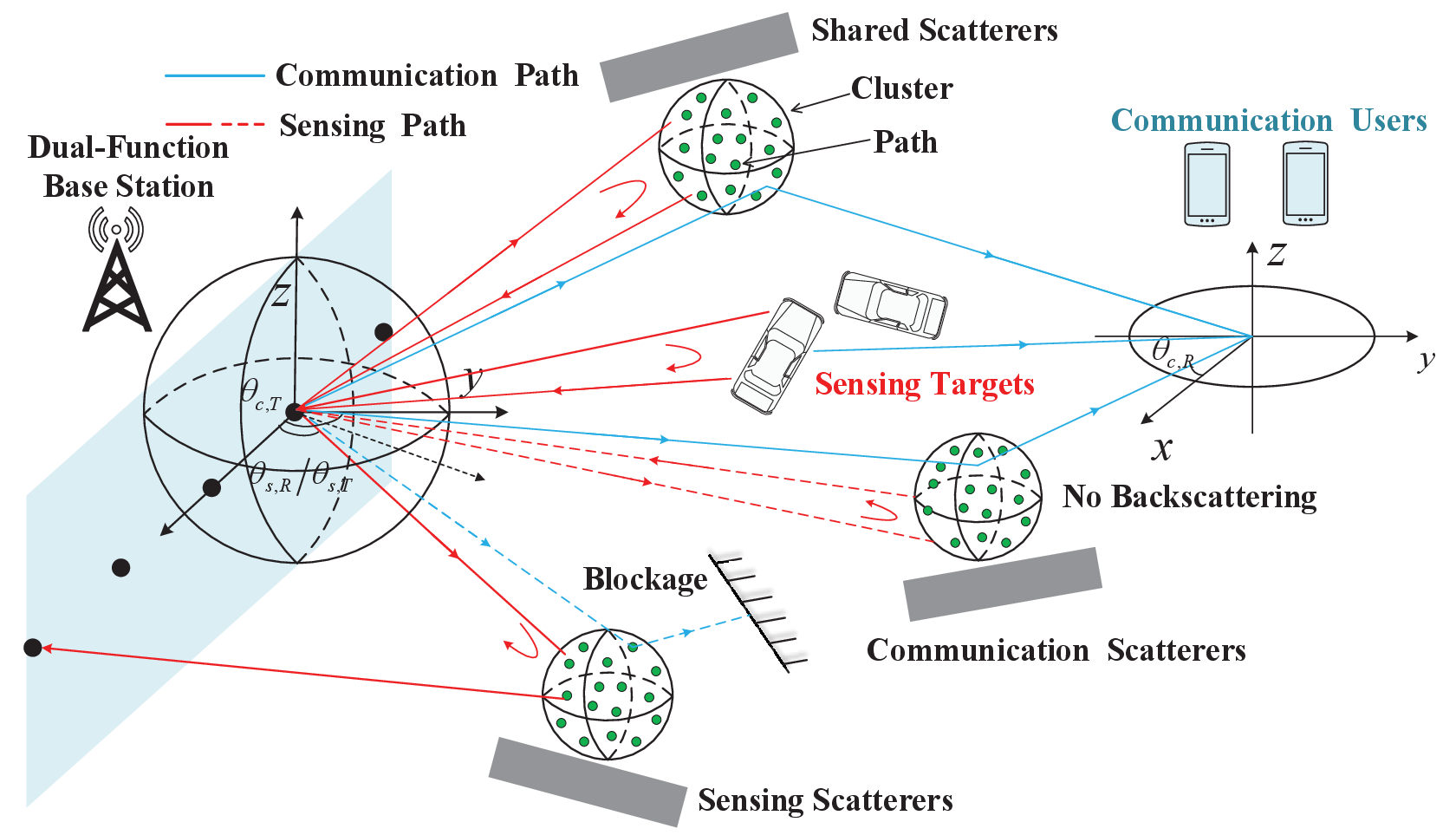}
\caption{Channel modeling for PMCW-NOMA ISAC systems.}
\label{fig1}
\end{figure}

\subsection{Channel Modeling for PMCW-NOMA ISAC}
As illustrated in Fig.~\ref{fig1}, the channel model incorporates paths related to sensing targets and communication users, which may be subject to environmental clutter and scattering.
\subsubsection{Sensing Channel Modeling}
According to the 3GPP ISAC channel modeling standardization \cite{ZhangISAC}, the sensing channel $\mathbf{H}^\text{s}\in {\mathbb{C}^{N_r \times N_t}}$ consists of two components: the target-related channel $\mathbf{H}_q\in {\mathbb{C}^{N_r \times N_t}}$ for sensing target $q(1\le q \le Q)$ and the background channel $\mathbf{H}_\text{b}\in {\mathbb{C}^{N_r \times N_t}}$ for environmental clutters. The round-trip channel $\mathbf{H}_{q}$ for target $q$ at slot $t$ can be expressed as
\begin{equation}
\mathbf{H}_{q}(t) = \alpha_q e^{j2\pi f_{d,q} t} [\mathbf{a}_{r}(\theta_{A,q})] [\mathbf{a}_{t}(\theta_{D,q})],
\end{equation}
where $\alpha_q$ is the complex path gain composed of the path loss and the radar cross section (RCS) of the target. $f_{d,q}$ is the Doppler shift associated with target $q$. Parameters $\theta_{D,q}$ and $\theta_{A,q}$ are the angles of departure and arrival for target $q$, respectively. $[\cdot]_k$ denotes the $k$-th element of a vector. The transmit and receive steering vectors $\mathbf{a}_{t}(\theta) \in \mathbb{C}^{N_t \times 1}$ and $\mathbf{a}_{r}(\theta) \in \mathbb{C}^{N_r \times 1}$ are given by
\begin{equation}
  \begin{aligned}
 &\mathbf{a}_{t}(\theta_{D,q}) = \left[1, e^{-j\frac{2\pi}{\lambda}d\sin\theta_{D,q}}, \dots, e^{-j\frac{2\pi}{\lambda}d(N_t-1)\sin\theta_{D,q}}\right]^T,\\
 &\mathbf{a}_{r}(\theta_{A,q}) = \left[1, e^{-j\frac{2\pi}{\lambda}d\sin\theta_{A,q}}, \dots, e^{-j\frac{2\pi}{\lambda}d(N_r-1)\sin\theta_{A,q}}\right]^T.
  \end{aligned}
\end{equation}

The total sensing channel $\mathbf{H}^\text{s}$ is the superposition of responses from all $Q$ targets and ${C^\text{b}}$ clustered clutters and each clutter consists of ${S^\text{b}_c}$ scatterers, which can be expressed as
\begin{equation}
\mathbf{H}^\text{s}(t) = \sum_{q=1}^{Q} \mathbf{H}_q(t) + \mathbf{H}_\text{b}(t),
\end{equation}
where the background channel $\mathbf{H}_\text{b}(t)$ is given by
\begin{equation}
\mathbf{H}_\text{b}(t) = \frac{1}{\sqrt{C^\text{b}S^\text{b}_c }}\sum\limits_{c = 1}^{C^\text{b}} \sum\limits_{s = 1}^{S^\text{b}_c}\alpha_{c,s} e^{j2\pi f_{d,{c,s}} t} \mathbf{a}_{r}(\theta_{A,{c,s}}) \mathbf{a}_{t}^T(\theta_{D,{c,s}}).
\end{equation}

\subsubsection{Communication Channel Modeling}
The channel vector from the $N_t$ transmit antennas to a single-antenna user $k$ is denoted by $\mathbf{h}_k^c(t) \in \mathbb{C}^{N_t \times 1}$. This channel is composed of paths reflected from the sensing targets and other environmental scatterers, which can be modeled as
\begin{equation}
\mathbf{h}_k^c(t) = \sum_{p=1}^{P} \beta_{k,p} e^{j2\pi f_{d,k,p} t} \mathbf{a}_{t}(\theta_{D,k,p})
\end{equation}
where $P$ denotes the all propagation paths, $\beta_{k,p}$, $f_{d,k,p}$, and $\theta_{D,k,p}$ are the complex gain, Doppler shift, and angle of departure for the $p$-th path to user $k$, respectively. 

\subsection{Signal Model for PMCW-NOMA ISAC}

\begin{figure}[t]
	\centerline{\includegraphics[width=3.5in]{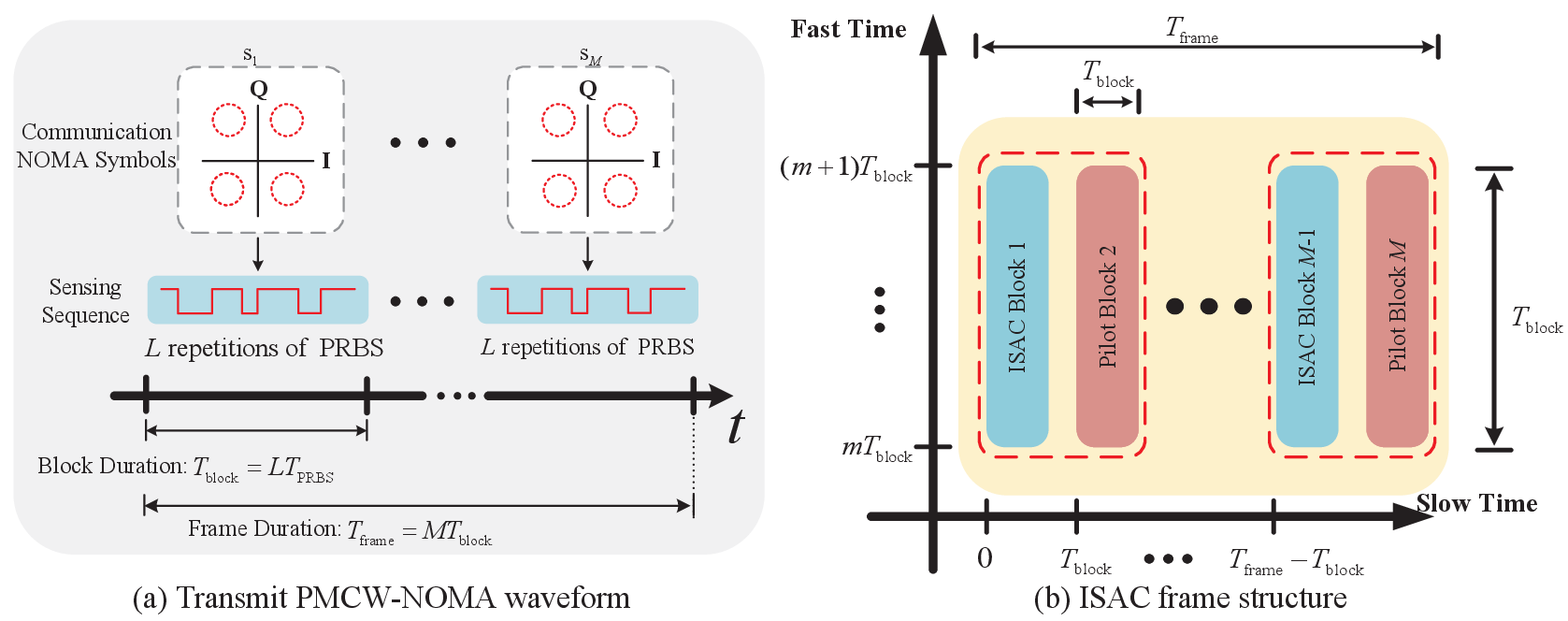}}
	\caption{{PMCW-NOMA waveform for ISAC.}}
	\label{fig2}
\end{figure}

\subsubsection{Transmitting Waveform} Fig.~\ref{fig2}(a) presents the transmitted PMCW-NOMA waveform frame at the DF-BS, where the transmitter sends $M$ PMCW blocks and the basic building code of a PMCW block is a PRBS with sequence length $L$, denoted as $\mathbf{c} \in \{-1, +1\}^{L \times 1}$. To enable MIMO operation, an orthogonal outer code is applied by using an $N_t \times N_t$ Hadamard matrix $\mathbf{W}$. The resulting outer-coded sequence matrix for all antennas is $\mathbf{P} = \mathbf{c} \otimes \mathbf{W} \in \mathbb{R}^{L N_t \times N_t}$, where $\otimes$ is the Kronecker product. 
In each PMCW block, the NOMA communication signal is carried. Considering a two-user NOMA system, let the far user be user 1 and the near user with superior channel condition be user 2. The transmitted symbol vector $\mathbf{s}_m \in \mathbb{C}^{N_t \times 1}$ during the $m$-th block ($1\le m \le M$) is a superposition of symbols for both users, which is given by
\begin{equation}
\mathbf{s}_{\text{noma},m} = \sqrt{p_1} \mathbf{s}_{1,m} + \sqrt{p_2} \mathbf{s}_{2,m},
\end{equation}
where $\mathbf{s}_{k,m}$ is the modulated symbol vector for user $k$, and $p_k$ is the NOMA power allocation coefficient, with $p_1+p_2=1$ and $p_1 > p_2$. 
The final transmitted signal matrix during $m$-th block is given by
\begin{equation}\label{waveform}
\mathbf{X}_m = \mathbf{P} \odot \left( \mathbf{1}_{L \cdot N_t} \mathbf{s}_{\text{noma},m}^T \right) \in \mathbb{C}^{(L \cdot N_t) \times N_t},
\end{equation}
where $\odot$ is the element-wise product and $\mathbf{1}_{L \cdot N_t}$ is a column vector of ones. 
\subsubsection{Sensing Signal Model}
The signal received by the $N_r$ antennas at the DF-BS is a superposition of reflections from all $Q$ targets, environmental clutter, and noise. For the $m$-th processing block, the received signal can be modeled as a data matrix $\mathbf{Y}_m \in \mathbb{C}^{N_r \times L N_t  }$, which can be expressed as
\begin{equation}
\mathbf{Y}_m =   \mathbf{H}^\text{s}_m\mathbf{X}_m^T + \mathbf{Z}_m,
\end{equation}
where $\mathbf{Z}_m\sim\mathcal{C}\mathcal{N}(0,{{{\sigma}_\text{rad}^{2} }\mathbf{I}_{N_r \times L N_t}})$ denotes complex Gaussian noise.
By combining the data $\mathbf{Y}_m$ from all $M$ blocks, we form a sensing data cube $\mathbf{Y}_{\text{rad}} \in \mathbb{C}^{ N_r \times LN_t \times M}$. 

\subsubsection{Communication Signal Model}
The baseband signal $\mathbf{y}_{k,m} \in {\mathbb{C}^{L N^t \times 1}}$ received at user $k$ during the $m$-th block is given by
\begin{equation}
\mathbf{y}_{k,m}  =  \mathbf{X}_m \mathbf{h}^{c}_{k,m} + \mathbf{z}_m,
\end{equation}
where $\mathbf{z}_m\sim\mathcal{C}\mathcal{N}(0,{{{\sigma}_\text{com}^{2} }\mathbf{I}_{L N_t}})$ is complex Gaussian noise. 

After the transmission of $M$ PMWC blocks, we can collect the overall received signal $\mathbf{Y}_{k} \in {\mathbb{C}^{L N^t \times M}}$ at user $k$. Then, the received signal is decoded with the Hadamard matrix $\mathbf{W}$ to yield data cube ${\mathbf{Y}}_{\text{dec},1} \in {\mathbb{C}^{L \times N^t \times M}}$ at the user $1$ and ${\mathbf{Y}}_{\text{dec}, 2} \in {\mathbb{C}^{L \times N^t \times M}}$ at user $2$ after temporal synchronization, respectively.
To realize the effective communication signal detection, the channel vector for each user $\mathbf{h}_k^c$ must first be estimated by transmitting the pilot signal. In Fig.~\ref{fig2}(b), the typical PMCW-NOMA ISAC frame structure are presented, where the dedicated pilot blocks are embedded into the ISAC frame to acquire the required channel state information. In the traditional NOMA detection processing, the far user treats the signal of the near user as noise to decoding the signal of the far user by the typical signal receiver, e.g., zero-forcing (ZF), while the near user utilize the SIC scheme to alleviate the inference of the far user. Note that the performance of SIC is limited by error propagation, where any errors in decoding the signal of far user are propagated and amplified when decoding the signal of near user. Moreover, traditional approaches rely on explicit pilot insertion, which consumes valuable spectral resources and reduces effective data throughput. However, since the sensing signal component of ISAC waveforms $\mathbf{X}$ inherently possess deterministic characteristics, we propose leveraging the outer-coded sensing sequence $\mathbf{P}$ in Eq.\eqref{waveform} as implicit pilots for channel estimation. By treating $\mathbf{P}$ as a structured training sequence, the proposed method eliminates the need for dedicated pilot overhead, which preserves spectral efficiency by repurposing existing sensing signals for dual use in channel estimation.

\section{Proposed Two-Timescale Transformer (T3former) for PMCW-NOMA ISAC Systems}
\begin{figure}[t]
	\centerline{\includegraphics[width=3.0in]{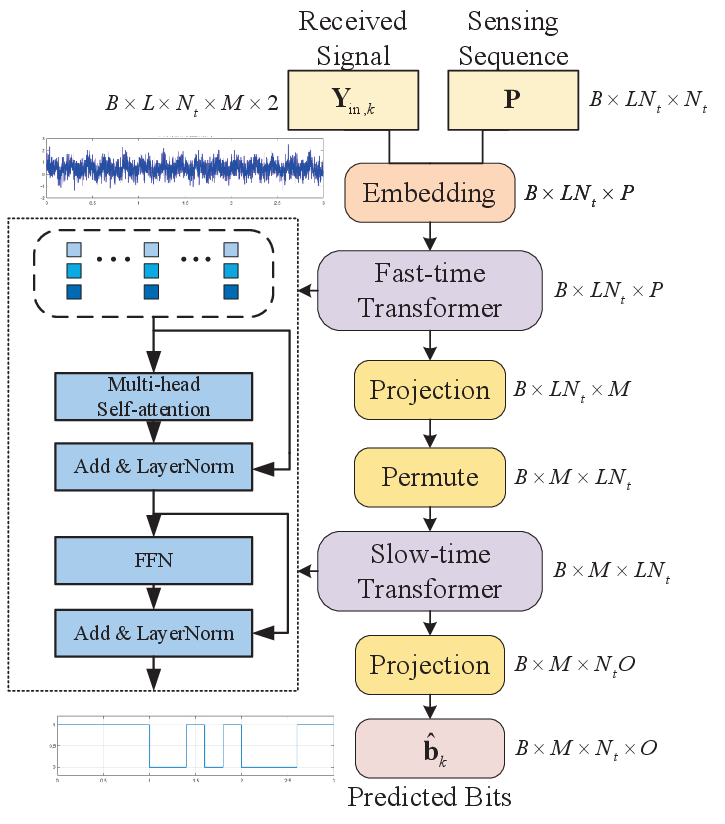}}
	\caption{{Proposed T3former for PMCW-NOMA ISAC systems.}}
	\label{fig4}
\end{figure}

In this work, we dispense with explicit channel estimation and equalization steps by developing an end-to-end network architecture. The goal is to directly map the received signal block $\mathbf{Y}_{\text{dec},k}$ and the known sensing sequence $\mathbf{P}$ to the transmitted bits for both users, which can be expressed as
\begin{equation}
f_\text{NN}: (\mathbf{Y}_{\text{dec},k} , \mathbf{P}) \rightarrow \{\hat{\mathbf{b}}_{k}\},
\end{equation}
where $\hat{\mathbf{b}}_{k}\in \mathbb{C}^{N_t \times M \times O}$ with modulation order $O$ are the estimated bit vectors by the proposed T3former model. 

\subsection{Dataset Construction and Preprocessing}
In the network training, the training dataset $\mathcal{D}$ consists of $N_s$ independent samples, where each sample $i$ is a tuple containing the received signal $\mathbf{Y}_{\text{dec},k}$, the deterministic sensing waveform $\mathbf{P}$, and the ground-truth transmitted bits $\mathbf{b}_k$. i.e., $
\mathcal{D} = \{ (\mathbf{Y}_{\text{dec},k,i}, \mathbf{P}, \mathbf{b}_{k,i}) \}_{i=1}^{N_s}.$
To adapt to the input of neural networks, the In-phase (I) and Quadrature (Q) components of the complex signal complex signal $\mathbf{Y}_{\text{dec},k,i}$ is split to form the input tensor ${\mathbf{Y}}_{\text{in},k} \in \mathbb{R}^{B \times L \times N_t \times M  \times 2}$ of the proposed T3former, where $B$ is the batch size in the network training. The raw input is first transformed into a sequence suitable for the Transformer, which involves reshaping, concatenation, and embedding.
First, the input tensor $\mathbf{Y}_{\text{in},k}$ is reshaped by combining the stream and I/Q dimensions. The sensing waveform $\mathbf{P}$ is similarly processed and broadcast to match the batch size. This results in two tensors: $\mathbf{Y}' \in \mathbb{R}^{B \times L \times N_t \times 2M}$ and $\mathbf{P}' \in \mathbb{R}^{B \times L \times N_t \times 2M}$, which are concatenated along the feature dimension, i.e., $
    \mathbf{Z}_0 = [\mathbf{Y}', \mathbf{P}'] \in \mathbb{R}^{B \times L \times N_t \times 4M}.$
    The tensor $\mathbf{Z}_0$ is reshaped into a tensor $\mathbf{Z}_1 \in \mathbb{R}^{B \times L_1 \times 4M}$, where $L_1 = L N_t$.
     In the feature embedding, a linear layer $\mathbf{W}_{\text{emb}}$ projects the input features to the embedding dimension of Transformer $D_{\text{h}}$, and positional encodings $\mathbf{E}_{\text{pos}}$ are added to provide the model with information about the sequence order, which is given by
    \begin{equation}
    \mathbf{X}_1 = \mathbf{Z}_1 \mathbf{W}_{\text{emb}} + \mathbf{E}_{\text{pos}} \in \mathbb{R}^{B \times L_1 \times D_{\text{h}}}.
    \end{equation}
    
\subsection{Fast-Timescale Feature Encoder $\mathcal{T}_1$}
To address the challenge of pilot-free signal detection, we propose an efficient T3former architecture in Fig.~\ref{fig4}, which jointly learns the channel characteristics and decode the superimposed NOMA signals directly from the received data. This model is built upon the popular Transformer network \cite{ref6}, which has demonstrated remarkable success in capturing long-range dependencies in sequential data. The architecture is composed of two sequential Transformer encoder stages, each operating on a different temporal and feature scale. 
In the proposed T3former, the first stage $\mathcal{T}_1$ processes the sequence $\mathbf{X}_1$. Each element in this sequence corresponds to a fast time block in Fig.~\ref{fig2}(b). This stage is designed to capture local features and correlations. $\mathcal{T}_1$ consists of $N_1$ identical Transformer encoder layers. A single layer is composed of the layer normalization (LN), the multi-head self-attention (MHSA) module and the feed forward network (FFN), which is defined by \cite{ref6}
\begin{subequations}
\begin{align}
   & \mathbf{A} = \text{LN}(\mathbf{X}_1), \\
    &\mathbf{A}' = \mathbf{A} + \text{MHSA}(\mathbf{A}), \\
   & \mathbf{B} = \text{LN}(\mathbf{A}'), \\
    &\mathbf{B}' = \mathbf{B} + \text{FFN}(\mathbf{B}).
\end{align}
 \end{subequations}
The MHSA is the core mechanism of the Transformer model, which is composed of multiple parallel scaled sot-product self-attention units. The self-attention operates on three inputs: a Query ($\mathbf{Q}$), a Key ($\mathbf{K}$), and a Value ($\mathbf{V}$). In the context of SA, $\mathbf{Q}$, $\mathbf{K}$, and $\mathbf{V}$ are all derived from the same input sequence, $\mathbf{A} \in \mathbb{R}^{L_{seq} \times D_{\text{h}}}$, where $L_{seq}$ is the sequence length.
First, the input $\mathbf{A}$ is projected into $\mathbf{Q}$, $\mathbf{K}$, and $\mathbf{V}$ using learned weight matrices $\mathbf{W}_Q, \mathbf{W}_K, \mathbf{W}_V \in \mathbb{R}^{D_{\text{h}} \times d_k}$:
\begin{subequations} \label{eq:25}
\begin{align}
\mathbf{Q} &= \mathbf{A} \mathbf{W}_Q,\\
\mathbf{K} &= \mathbf{A} \mathbf{W}_K, \\
\mathbf{V} &= \mathbf{A} \mathbf{W}_V.
\end{align}
\end{subequations}
Furthermore, the scaled dot-product SA is given by
\begin{equation}
\text{Attention}(\mathbf{Q}, \mathbf{K}, \mathbf{V}) = \text{softmax}\left(\frac{\mathbf{Q}\mathbf{K}^T}{\sqrt{d_k}}\right)\mathbf{V},
\end{equation}
where $\mathbf{Q}\mathbf{K}^T$ computes the similarity scores between each query and all keys. $d_k$ is the dimension of the key vectors.
The $\text{softmax}$ function normalizes the scores into attention weights, which sum to 1. The result is multiplied by $\mathbf{V}$ to produce the final output, where values corresponding to high attention weights are emphasized.

The output of $\mathcal{T}_1$ is a sequence of hidden representations $\mathbf{H}_1 = \mathcal{T}_1(\mathbf{X}_1) \in \mathbb{R}^{B \times L_1 \times D_{\text{h}}}.$, which has the same dimensions as the input $\mathbf{X}_1$.
Finally, the fully connected (FC) layer of this encoder projects the hidden dimension to match the number of PMCW blocks $M$, i.e., $\mathbf{H}'_1 = \text{FC}_1(\mathbf{H}_1) \in \mathbb{R}^{B \times L_1 \times M}.$

\subsection{Slow-Timescale Feature Encoder $\mathcal{T}_2$}
The slow-timescale encoder $\mathcal{T}_2$ in Stage 2 is designed to aggregate the local features learned in Stage 1 into a global context, where the feature dimensions are permuted to obtain $\mathbf{X}_2 = \text{Permute}(\mathbf{H}'_1) \in \mathbb{R}^{B \times M \times L_1}$. The new input sequence $\mathbf{X}_2$ has a length of $L_2=M$. Each element in this sequence is a vector representing the aggregated features for a single stream across all PMCW blocks. The second encoder $\mathcal{T}_2$ consisting of $N_2$ Transformer layers is applied to $\mathbf{X}_2$, which is given by
    \begin{equation}
    \mathbf{H}_2 = \mathcal{T}_2(\mathbf{X}_2) \in \mathbb{R}^{B \times S \times L_1}.
    \end{equation}
    
    The second encoder operates on the slow-time scale, as it processes the sequence of streams. It allows the model to learn cross-stream interference and aggregate features that are distributed globally across the entire PMCW blocks.  Finally, a FC-based linear output head projects the globally-aware hidden states $\mathbf{H}_2$ to the desired output dimension, followed by a final reshape to produce the estimated bits
$\hat{\mathbf{b}}_k = \text{Reshape}(\text{FC}_2(\mathbf{H}_2)) \in \mathbb{R}^{B \times M \times N_t \times O}$.
This two-timescale design enables the model to build a hierarchical understanding of the signal, moving from local channel effects to global signal structure, which is crucial for robust detection in a pilot-free environment. 

The model is trained end-to-end in a supervised manner. Since the task is to recover binary data, we treat it as a binary classification problem for each transmitted bit. The output of the model is a tensor of logits. We use the binary cross-entropy (BCE) with logits loss function, which combines a sigmoid layer and a BCE loss in one single class for better numerical stability. The loss function for a single sample is given by
\begin{equation}
\mathcal{L}(\mathbf{b}, \hat{\mathbf{b}}) = -\frac{1}{N_{b}} \sum_{i=1}^{N_{b}} [b_i \cdot \log(\sigma(\hat{b}_i)) + (1-b_i) \cdot \log(1-\sigma(\hat{b}_i))],
\end{equation}
where $N_{b}$ is the total number of bits in the output, $b_i$ is the true value of the $i$-th bit, $\hat{b}_i$ is the corresponding logit predicted by the model, and $\sigma(\cdot)$ is the sigmoid function. By training under different datasets, the proposed T3former is deployed at both near and far user terminals.

\section{Numerical Results}
\begin{figure}[!t]
\centering
\includegraphics[width=2.5in]{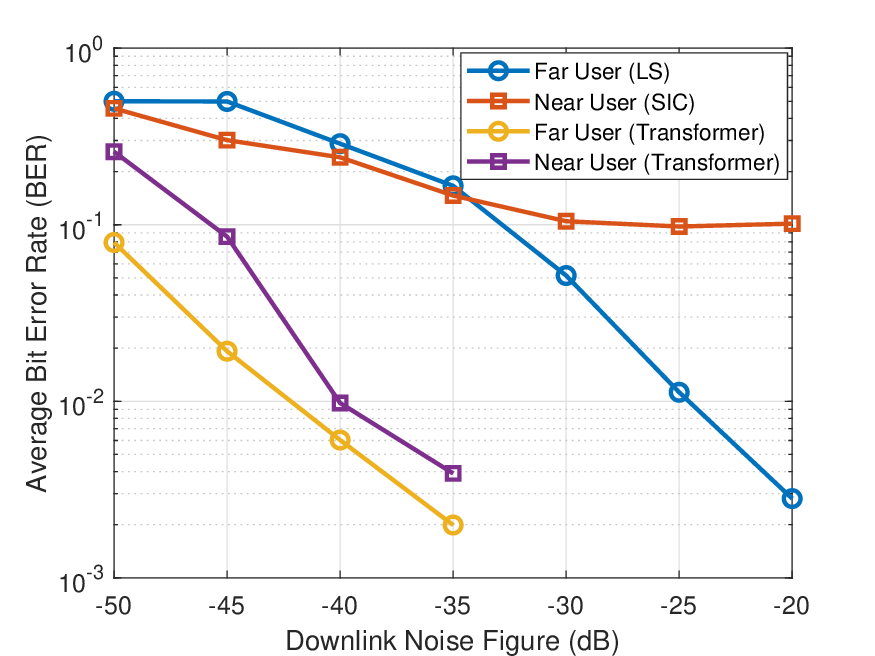}
\caption{BER vs. downlink noise figure for different schemes.}
\label{fig_ber}
\end{figure}

In the simulation, we consider a typical mmWave ISAC scenario, Table \ref{tab:params} summarize the simulation parameters of the considered PMCW-NOMA ISAC system and the proposed T3former. 
\begin{table}[!t]
\caption{Simulation Parameters}
\centering
\begin{tabular}{|l|l|}
\hline
\multicolumn{2}{|c|}{\textbf{System Parameters of PMCW-NOMA ISAC }} \\
\hline
Carrier frequency & 77 GHz \\
Bandwidth & 150 MHz \\
Transmit/receive antennas ($N_t, N_r$) & (16, 16) \\
Locations of users ($U_1, U_2$) & ([15,15,0],[5,5,0]) \\
Sensing range ($r, \theta$) & ($[0,60], [-\pi/2,\pi/2]$) \\
Number of target ($Q$) & 2 \\
Moving speed of targets ($v_1,v_2$) & ([4,4,2],[5,5,0])m/s \\
PMCW code length ($L$) & 63 \\
Number of PMCW blocks ($M$) & 8 \\
Sensing block periodicity ($M_p$) & 4 \\
Modulation scheme & QPSK ($O=4$) \\
NOMA power allocation ratio ($p_1, p_2$) & 0.7 , 0.3 \\
\hline
\multicolumn{2}{|c|}{\textbf{Hyper-parameters of T3former}} \\
\hline
Optimizer & Adam \\
Number of samples  $N_s$ & $20000$ \\
Learning rate with Cosine decay & $1 \times 10^{-4}$  \\
Batchsize & 16 \\
Number of epochs & 100 \\
Dimension of embedding tensor $D_{\text{h}}$ & 256\\
Dimension of Key tensor $d_k$ & 64\\
Number of layers $N_1$ in encoder $\mathcal{T}_1$  & 3\\
Number of layers $N_2$ in encoder $\mathcal{T}_2$  & 6\\

\hline
\end{tabular}
\label{tab:params}
\end{table}
Fig. \ref{fig_ber} shows the BER versus the downlink noise figure of different signal detection schemes for both the far user and the near user. It is evident that the proposed T3former-based receiver significantly outperforms the traditional ZF/SIC schemes for both users across the entire range of SNRs. For the near user, the performance gain is particularly pronounced. The performance of traditional SIC receiver is limited by error propagation. In contrast, the proposed T3former performs joint detection, learning to effectively cancel the interference without explicit and error-prone subtraction. For the far user, the Transformer model demonstrates a superior ability to learn and equalize the channel from the implicit pilots compared to the linear ZF baseline.

\begin{figure}[!t]
\centering
\includegraphics[width=2.5in]{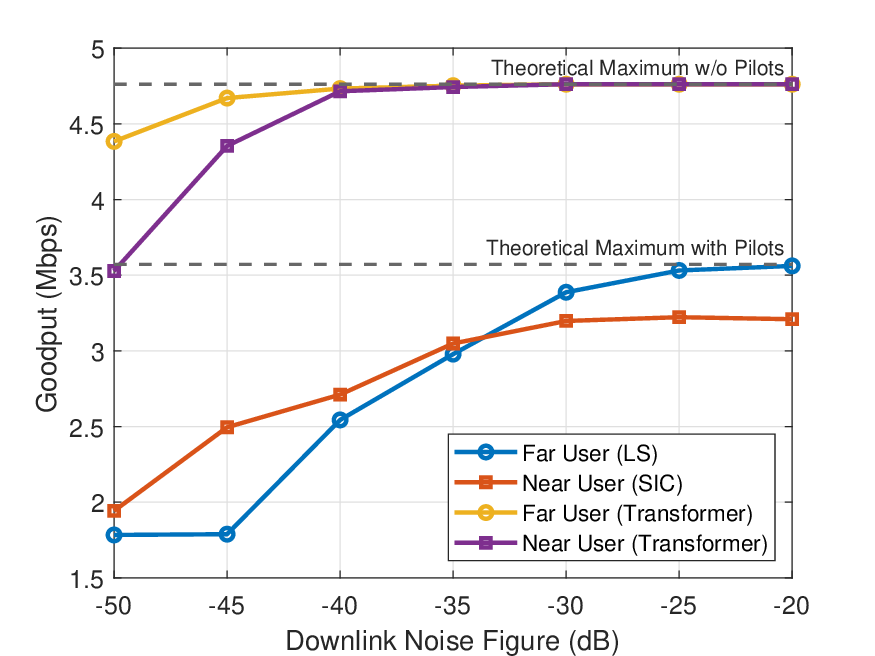}
\caption{Goodput vs. downlink noise figure for different schemes.}
\label{fig_goodput}
\end{figure}

Fig. \ref{fig_goodput} presents the system Goodput of different signal detection schemes. Goodput is calculated as $G =R_\text{max} \times (1 - \text{BER})$, where $R_\text{max}=\frac{(M-M/M_p)\times N_t \times O}{M\times T_\text{block}}$ is the theoretical maximum data rate and $T_\text{block}$ denotes the block duration. Here, $M/M_p=2$ for traditional schemes with pilot blocks while $M/M_p=0$ for the proposed pilot-free system. Due to its lower BER, the Goodput of the proposed T3former is substantially higher than ZF/SIC methods at all SNR levels. Second, the proposed pilot-free design raises the ceiling on achievable throughput, where the proposed T3former-based receiver rapidly approaches this higher theoretical maximum that the pilot-free system utilizes all blocks for data transmission. 

\begin{figure}[!t]
\centering
\subfloat[\scriptsize
 Range-Angle response.]{ \includegraphics[width=1.7in]{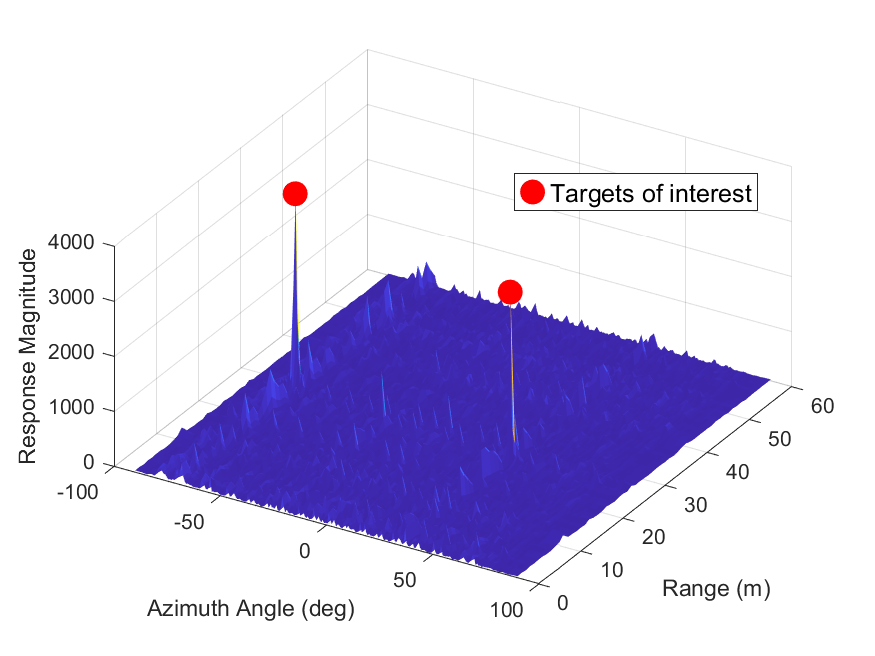}%
\label{fig_first_case}}
\hfil
\subfloat[\scriptsize
 Range-Doppler response.]{\includegraphics[width=1.7in]{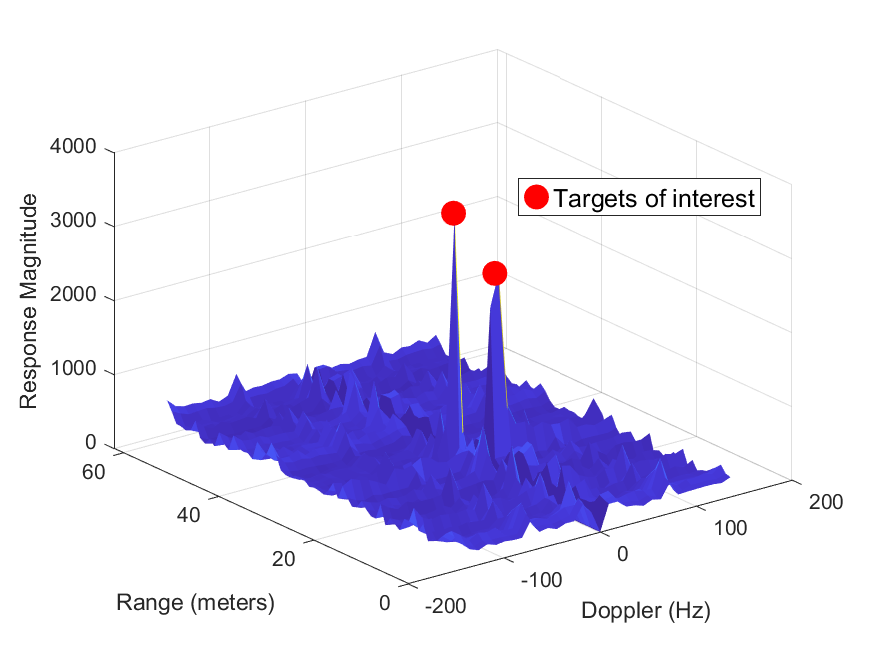}%
\label{fig_second_case}}
\caption{Sensing performance of the ISAC system.}
\label{fig_sensing}
\end{figure}

Fig. \ref{fig_sensing} provides the sensing results obtained from the ISAC signal processing chain by the match filtering method \cite{Sturm2011Waveform}, which demonstrate the system sensing capabilities are not compromised under the designed pilot-free waveform. The range-angle response in Fig. \ref{fig_sensing}(a) and range-Doppler response in Fig. \ref{fig_sensing}(b) both show distinct and sharp peaks corresponding to the true locations and velocities of the simulated targets. This confirms that the PMCW-MIMO waveform, even when modulated with NOMA communication data, retains its excellent sensing properties, validating that the proposed pilot-free ISAC system achieves both of its objectives effectively.

\section{Conclusion}
In this paper, we have proposed a pilot-free PMCW-NOMA ISAC system to enhance communication throughput. By leveraging the inherent structure of the PMCW waveform as an implicit pilot, the proposed system eliminates the resource overhead of traditional pilot-based schemes. To address the complex task of joint channel estimation and multi-user detection in this pilot-free and interference-rich environment, we developed a T3former-based receiver architecture that demonstrates a powerful ability to learn the intricate mappings from the received signal to the transmitted bits. Numerical results have shown that the proposed deep learning approach yields significant performance gains over traditional methods, achieving lower BER and substantially higher Goodput. Future work could explore the model adaptability to more complex and dynamic channel environments and extend the framework to support different ISAC waveforms.

\end{document}